\documentclass{article}

\usepackage{PRIMEarxiv}

\usepackage[utf8]{inputenc} 
\usepackage[T1]{fontenc}    
\usepackage{hyperref}       
\usepackage{url}            
\usepackage{booktabs}       
\usepackage{amsmath,amssymb,amsfonts}      
\usepackage{nicefrac}       
\usepackage{microtype}      
\usepackage{lipsum}
\usepackage{graphicx}
\graphicspath{ {./images/} }
\usepackage{cite}
\usepackage{doi}
\usepackage{float} 
\usepackage{mathrsfs}
\usepackage{subfigure}
\usepackage{threeparttable}
\usepackage{multicol}
\usepackage{multirow}
\usepackage{tabto}
\usepackage{algorithm}
\usepackage{algpseudocode}
\usepackage{makecell}
\usepackage{array}
\usepackage{diagbox}

\title{Temporal Evolution of Knee Osteoarthritis: A Diffusion-based Morphing Model for X-Ray Medical Image Synthesis
}

\author{
  Zhe Wang, Mohamed Jarraya$\dagger$ \\
  Department of Radiology\\ Massachusetts General Hospital\\
  Harvard Medical School \\
  Boston, 02114, USA\\
  \texttt{$\{$zwang78, mjarraya$\}$@mgh.harvard.edu} \\
   \And
  Aladine Chetouani \\
  PRISME Laboratory, EA 4229\\
  University of Orleans\\
  Orleans, 45067, France \\
  \texttt{aladine.chetouani@univ-orleans.fr} \\
    \And
  Rachid Jennane \\
  IDP Institute, UMR CNRS 7013\\ University of Orleans\\
  Orleans, 45067, France \\
  \texttt{rachid.jennane@univ-orleans.fr} \\
      \And
  Yuhua Ru \\
  Jiangsu Institute of Hematology\\  First Affiliated Hospital of Soochow University\\
  Suzhou, 215006, China\\
  \texttt{ruyuhua@163.com} \\
  \And
  Wasim Issa \\
  Department of Orthopedic Oncology\\ Cancer Hospital of Pernambuco\\
  Recife, 50040, Brazil\\
  \texttt{drwasimissa@gmail.com} \\
}

\begin{document}
\maketitle

\begin{abstract}
Knee Osteoarthritis (KOA) is a common musculoskeletal disorder that significantly affects the mobility of older adults. In the medical domain, images containing temporal data are frequently utilized to study temporal dynamics and statistically monitor disease progression. While deep learning-based generative models for natural images have been widely researched, there are comparatively few methods available for synthesizing temporal knee X-rays. In this work, we introduce a novel deep-learning model designed to synthesize intermediate X-ray images between a specific patient's healthy knee and severe KOA stages. During the testing phase, based on a healthy knee X-ray, the proposed model can produce a continuous and effective sequence of KOA X-ray images with varying degrees of severity. Specifically, we introduce a Diffusion-based Morphing Model by modifying the Denoising Diffusion Probabilistic Model. Our approach integrates diffusion and morphing modules, enabling the model to capture spatial morphing details between source and target knee X-ray images and synthesize intermediate frames along a geodesic path. A hybrid loss consisting of diffusion loss, morphing loss, and supervision loss was employed. We demonstrate that our proposed approach achieves the highest temporal frame synthesis performance, effectively augmenting data for classification models and simulating the progression of KOA.
\end{abstract}


\section{Introduction}
Knee OsteoArthritis (KOA) is a progressive disease marked by the degeneration and damage of articular cartilage, changes at the joint margins, and reactive hyperplasia of the subchondral bone \cite{kneeoa}. Various factors, such as age, obesity, stress, and trauma, can contribute to its development \cite{multi-factor}. Those affected often endure severe pain and limited mobility, which can greatly diminish their quality of life and elevate the risk of chronic conditions like cardiovascular disease \cite{cardiovascular}. Despite extensive research, the precise cause of KOA remains unknown, and there is currently no cure \cite{notclear}. Hence, early detection is crucial for implementing timely interventions, such as weight reduction, to delay the onset and progression of KOA \cite{weightloss}.

In 1957, Kellgren and Lawrence introduced the Kellgren-Lawrence (KL) grading system for assessing KOA \cite{KL}. According to Table \ref{KL_grades}, this system categorizes KOA into five distinct grades, determined by the presence and severity of symptoms such as osteophytes and Joint Space Narrowing (JSN). The KL grading system, despite being widely adopted, serves as a semi-quantitative method. Therefore, the diagnosis relies heavily on the practitioner's experience and subjective judgment, which can lead to varying grades being assigned to the same knee X-ray image by different medical professionals \cite{shamir}.

\begin{table}[htbp]
\centering
\caption{Detailed Overview of the KL Grading System}
\setlength{\tabcolsep}{1.3mm}
\begin{tabular}{lll}
\toprule
Grade & Severity & Detailed description \\
\midrule
KL-0 & None & Definitive absence of any osteoarthritis signs \\
KL-1 & Doubtful & Possible presence of initial osteophytic lipping \\
KL-2 & Minimal & Certain osteophytes formation and potential JSN \\
KL-3 & Moderate & Multiple moderate osteophytes, confirmed JSN, some\\
&& bone sclerosis, and potential bone end deformities \\
KL-4 & Severe & Large and numerous osteophytes, confirmed JSN, and\\
&& definitive deformation of bone ends \\
\bottomrule
\end{tabular}
\label{KL_grades}
\end{table}

Exploring the progression of pathological changes is one of the crucial tasks in medical imaging diagnosis and planning disease treatment \cite{lambin2017radiomics}. Particularly for KOA, which has limited effective data and is incurable in its late stages \cite{hu2022adversarial}, generating X-ray images representing the continuous severity levels of KOA not only can simulate the progression of the disease, making the observation of its development more intuitive, but also can serve as a data augmentation technique. Especially for the KL-1 grade, which lies between healthy (KL-0) and confirmed KOA (KL-2) and is reported as doubtful KOA \cite{favero2015early}, its diagnosis presents certain challenges. During this transitional stage, the X-ray images may show subtle changes, which might be difficult to detect for medical professionals without extensive experience, yet are crucial for early diagnosis and intervention \cite{mahmoudian2021early}.

With the rapid advancements in deep learning technologies, the field of medical image generation has seen significant growth \cite{greenspan2016guest}, particularly with the Generative Adversarial Networks (GANs) \cite{GAN} and the Denoising Diffusion Probabilistic Model (DDPM) \cite{ddpm}. Despite their success, GAN-based models often face challenges related to training stability, including mode collapse and the complex task of finding a Nash equilibrium between the generator and discriminator \cite{kossale2022mode}. This instability can hinder the model's ability to produce diverse and realistic images, limiting its applicability in critical medical image generation tasks \cite{beat}. Conversely, the training process of DDPM, characterized by its iterative refinement of generated images through successive diffusion steps, inherently promotes the creation of high-quality and diverse samples \cite{guo2023diffusion}, which not only mitigates the risk of generating images with artefacts or unrealistic features but also aligns closely with the nuances and complexity of medical imaging data \cite{deshpande2024assessing}.

Several diffusion-based learning models have been developed. In \cite{DDPM_Packh}, Packhauser et al. employed a latent diffusion model \cite{latent_diffusion_model} to generate high-quality class-conditional chest X-ray images. They introduced a sampling strategy designed to preserve the privacy of sensitive biometric information during image generation. The results demonstrated that their approach surpassed GAN-based methods. In \cite{pinaya}, Pinaya et al. introduced a swift method utilizing DDPM for the detection and segmentation of anomalous regions in brain MRI scans. Their approach involves generating a healthy reference sample and identifying the anomaly by subtracting this generated sample from the input image to create the segmentation map. In \cite{diffcmr}, Xiang et al. emphasized leveraging conditional denoising diffusion probabilistic models for reconstructing MRI images from under-sampled data, highlighting the advantages of these models in terms of training and tuning compared to previous GAN-based methods. In \cite{cola}, Jiang et al. introduced the pioneering diffusion-based model for multi-modality MRI synthesis, which features an architecture that minimizes memory usage by functioning within the latent space. In \cite{3D_generation}, Khader et al. explored the application of diffusion probabilistic models for synthesizing high-quality 3D medical images, specifically in MRI and CT imaging. The study evaluated synthetic image quality through expert radiologist assessment and demonstrated that these models can produce realistic images. It highlighted the potential of using synthetic images in self-supervised pre-training to improve the performance of breast segmentation models, especially in situations with limited data availability.

Learning-based morphing image registration methods have been developed due to their capability to provide morphing fields in real-time, allowing the source to be transformed into the target \cite{review_field}. Smooth morphing fields enable the source to morph with topology preservation \cite{montagnat2001review}. Using this property, the morphed knee image might not only meet the KOA progression but also maintain all the original texture details. Drawing inspiration from the method by which DDPM produces images using the latent space derived from the parameterized Gaussian process, a novel Diffusion-based Morphing Model (DMM) to visualize the temporal evolution of KOA was proposed in this study. Specifically, our proposed approach was composed of diffusion and morphing modules. Given a source image and a target image, the latent code was learned using a score function of the DDPM. Then, this latent code was fed into the morphing module to synthesize intermediate frames through the morphing fields in an image registration manner, ensuring the topology of the source image was preserved. Moreover, to ensure that the synthesized intermediate frames corresponded more accurately to the respective KOA grade, a supervisor (i.e., a pre-trained classification model for KL-2 vs. KL-3) was introduced. Once the model was trained, continuous KOA X-rays could be synthesized from healthy knee X-ray images by averaging the morphing fields (i.e., from KL-0 to KL-1, KL-2, KL-3, and KL-4). Experimental results demonstrated that the synthesized X-ray images were valid as they effectively reflected the progression of KOA while maintaining the structural and textural integrity of the original images.


The primary contributions of this paper include the following:
\begin{itemize}
\item[$\bullet$] A DMM is proposed for the continuous synthesis of knee X-ray images.
\item[$\bullet$] The synthesized intermediate frames are visualized to observe the progression of KOA.
\item[$\bullet$] The generalization validation of the proposed approach is employed using a multi-faceted approach.
\item[$\bullet$] All experimental results are sourced from the OsteoArthritis Initiative (OAI) \cite{OAI} database.
\end{itemize}

\section{Proposed Approach}
\label{proposed_approach}
In this section, the classical DDPM model, our proposed model, and the employed hybrid loss strategy are presented.

\subsection{Classical DDPM model}
Before delving into the proposed approach, we provide a brief overview of the DDPM, which serves as the foundation for our methodology.
\label{classical introduction}

\begin{figure}[htbp]
\centering
\includegraphics[width=0.75\textwidth]{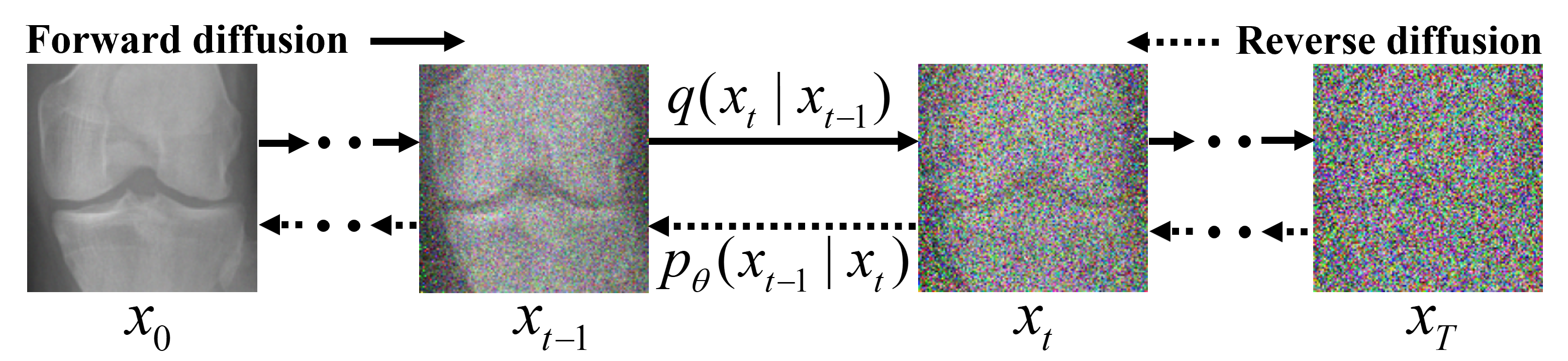}
\caption{The structure of the classical DDPM model. The sequence from left to right shows the original X-ray image, followed by ones with increasing levels of added noise.}
\label{ddpm}
\end{figure}

As shown in Fig. \ref{ddpm}, DDPM is a generative model that aims to synthesize samples from noisy data by learning the underlying probability distribution of the clean data. At its core, the concept of DDPM revolves around modeling the evolution of data as it undergoes a diffusion process. This process involves gradually adding noise to an initial data point and then attempting to recover the original signal through denoising. By iteratively applying noise and denoising, DDPM captures the underlying probability distribution of the data, allowing them to synthesize realistic samples \cite{nichol2021improved}.

\subsubsection{Diffusion process}
A diffusion process describes the evolution of a random variable over time. Through the forward diffusion process, for the original data $x_0 \sim q(x_0)$, each step of the diffusion process, which comprises a total of $T$ steps, involves adding Gaussian noise $n$ to the data obtained from the previous step $x_{t-1}$ as follows:

\begin{equation}
q(x_t\mid x_{t-1})=\mathcal{N} (x_t;\sqrt[]{1-\beta_t}x_{t-1},\beta_t\mathbf{I})
\end{equation}
where $\beta_t$ represents the variance used for each step in the range of $[0,1]$, and $\mathbf{I}$ represents the identity matrix. The entire diffusion process constitutes a Markov chain \cite{norris1998markov}:

\begin{equation}
q(x_{1
}\mid x_0)=\prod_{t=1}^{T} q(x_t\mid x_{t-1})
\end{equation}

Here, we directly sample $x_t$ for any step $t$ based on the original data $x_0$: $x_t \sim q(x_t \mid x_0)$ defining $\alpha_t = 1-\beta_t$ and $\bar{\alpha}t = \prod{i=1}^{t} \alpha_i$. Through the reparameterization technique, we have:

\begin{equation}
\begin{aligned}
x_t &= \sqrt{\alpha_t}x_{t-1} + \sqrt{1-\alpha_t}n_{t-1} \\&= \sqrt  {\alpha _ {t}} (\sqrt  {  \alpha _ {t-1}}  x_ {t-2}  +  \sqrt {1-\alpha _ {t-1}}  n_{t-2}) +  \sqrt {1-\alpha _ {t}}n_{t-1}\\&=\sqrt{\bar{\alpha}_t}x_0 + \sqrt{1-\bar{\alpha}_t}n, \quad \quad \forall n_t \sim \mathcal{N}(0, \mathbf{I})
\end{aligned}
\label{x_t}
\end{equation}

\subsubsection{Denoising and Sampling}
After the diffusion process, denoising is performed by sampling from the updated noise distribution. Estimating distribution $q(x_{t-1}\mid x_t)$ requires the utilization of the entire training set. Typically, a neural network, such as a U-Net \cite{ronneberger2015u}, is employed to estimate these distributions. Here, the reverse process is also defined as a Markov chain composed of a sequence of Gaussian distributions parameterized by neural network parameters:

\begin{equation}
p_\theta(x_{0:T}) = p(x_T)\prod_{t=1}^{T} p_\theta(x_{t-1}\mid x_t)
\end{equation}

\begin{equation}
p_\theta(x_{t-1}\mid x_t)=\mathcal{N} (x_{t-1};\mu_\theta(x_t,t), \Sigma_{\theta}(x_t,t))
\end{equation}
where $p(x_T)=\mathcal{N} (x_T;0,\mathbf{I})$, and $p_\theta(x_{t-1}\mid x_t)$ represent parameterized Gaussian distributions, with their mean $\mu_\theta(x_t,t)$ and variance $\Sigma_{\theta}(x_t,t)$ determined by trained learning networks.

\begin{figure*}
\centering 
\includegraphics[width=1\textwidth]{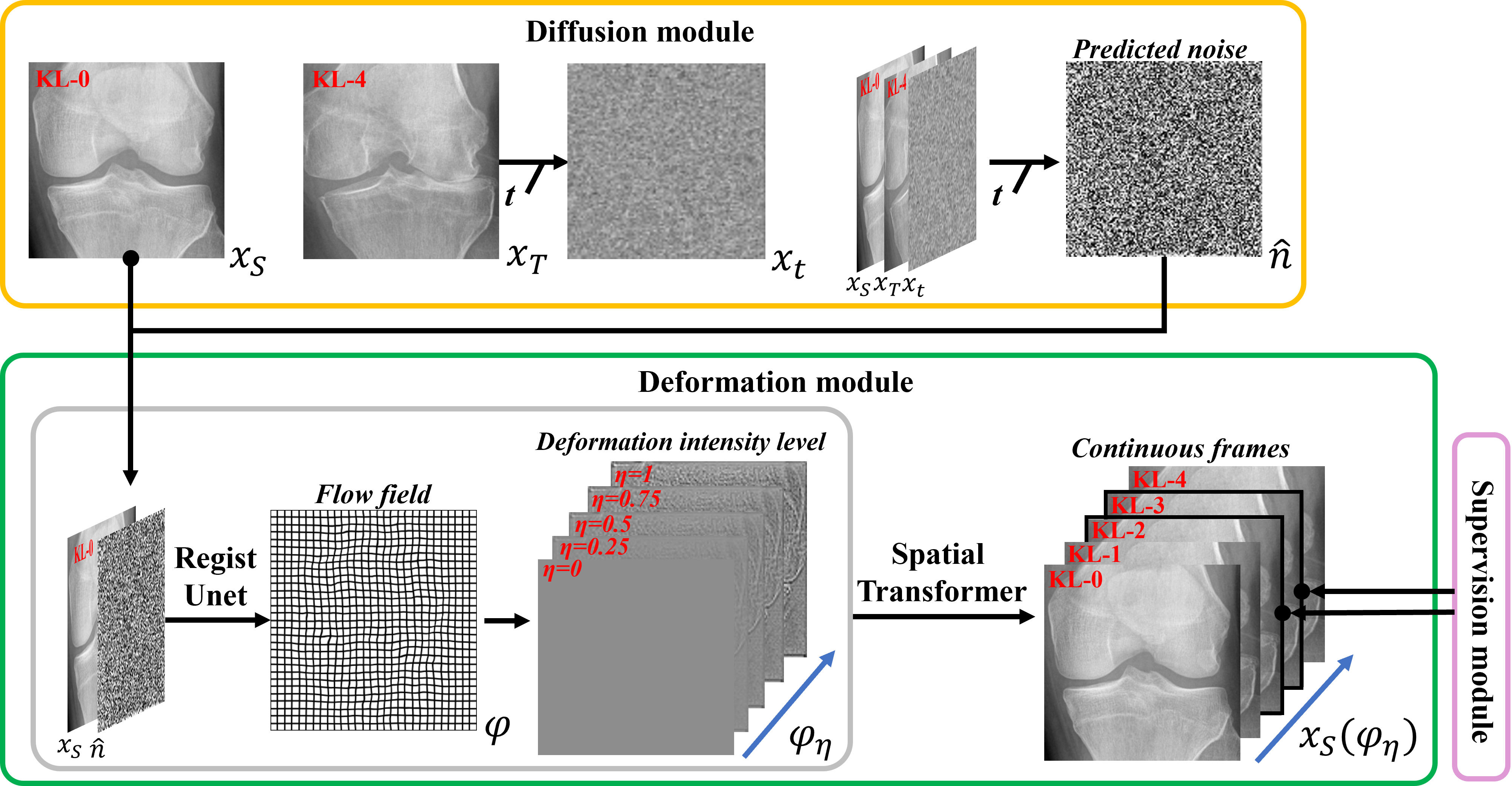}
\caption{Flowchart of the proposed approach, with black arrows representing the data flow. First, the source $x_S$ and target $x_T$ knee X-ray images are fed into the diffusion module, where the source image $x_T$ undergoes a process that adds noise to the image, incrementally increasing its noise level. Then, the source $x_S$, target $x_T$, and noised $x_t$ images are utilized to predict the noise $\hat{n}$. Following the diffusion module, the predicted noise $\hat{n}$ and the source $x_S$ image are processed through the morphing module, which applies transformations to the images via a flow field $\varphi$ to simulate structural changes. The different levels of morphing are donated by $\eta$ values. After morphing, the synthesized images $x_S{(\varphi_\eta)}$ are supervised by a supervision module to ensure they more accurately correspond to each stage of KOA. Further details are provided in Section \ref{proposed_model}.}
\label{flowchart}
\end{figure*}

\subsubsection{Training process}
The parameters $\theta$ of the diffusion module aim to minimize the distance between the true noise $n$ and the predicted noise $\hat{n}$. The loss function quantifies this difference as the expected value of the squared norm between the actual noise and the noise estimated by the neural network.

\begin{equation}
\mathcal{L}_ {diff} =  \mathbb{E}_ {x_0, n \sim \mathcal{N}(0,\mathbf{I})} \left [|| n -  \hat{n}  ( \sqrt {\bar \alpha _ {t}}  x_ {0}  +  \sqrt {1-\bar\alpha _ {t}}n ,t)  ||^ {2} \right ]
\label{diff_classical_loss}
\end{equation}
where $\mathbb{E}$ represents the expectation over the joint distribution of the original data $x_0$ and Gaussian noise $n$.

\subsection{Proposed approach}
\label{proposed_model}

As presented in Fig. \ref{flowchart}, our approach consists of three main modules: a diffusion module, a morphing module, and a supervisor module. Source images $x_S$ and target images $x_T$ are a set of KL-0 and KL-4 knee X-ray images from the same individuals. Since our objective is to generate continuous frames from KL-0 image, we here define the perturbed target $x_t$ denoting $x_0 = x_T$ according to Eq. \ref{x_t}:

\begin{equation}
x_t = \sqrt{\bar{\alpha_t}}x_T + \sqrt{1-\bar{\alpha_t}}n
\end{equation}

Leveraging the condition and the combination of the perturbed target $x_t$, the source image $x_S$, and the target one $x_T$, the latent code $c$ is generated using the designed U-Net during the reverse diffusion. The employed 2D U-Net architecture follows a multi-scale hierarchical structure. It consists of a time embedding module, an encode module, a middle module, and a decode module. The encoder module begins with a 2D convolutional layer using a kernel size of 3 and padding of 1, followed by a series of residual blocks with increasing feature dimensions. Each residual block contains multiple convolutional layers, GroupNorm \cite{wu2018group} for normalization, Swish activation function for non-linearity, and optional dropout for regularization. The middle module introduces self-attention modules that enable the capture of long-range dependencies in the data. Throughout the encoder, downsampling is achieved using 2D convolutional layers with a kernel size of 3, stride of 2, and padding of 1, which effectively reduces the spatial dimensions while increasing the number of feature channels. In the decoder, residual blocks are employed again. Upsampling is performed through a combination of 2D convolutional transpose layers with a kernel size of 2, stride of 2, and padding of 0, followed by residual blocks to fuse features from the encoder and decoder pathways.

Then, the source image $x_S$ and predicted noise $\hat{n}$ are fed to a registration network to obtain a morphing flow field $\varphi$. The employed registration network follows a U-Net-like structure. The input block performs an initial convolution with downsampling to capture low-level features. Subsequently, a contracting path is formed using downsampling blocks that progressively reduce spatial dimensions while increasing channel complexity. On the other hand, the expansive path employs upsampling blocks to recover spatial details and enhance features. These upsampled features are then concatenated with features from the contracting path, facilitating information fusion. Before generating the final output, a refinement block refines the concatenated features. Finally, the output block generates the predicted morphing flow field $\varphi$. It is noteworthy that here we introduce a scale parameter $\eta$ for the flow field $\varphi$, which represents five uniformly distributed numbers within the range $[0, 1]$ as the adjustable morphing intensity level (i.e. $\varphi_\eta = \eta \varphi$) for the synthesized intermediate frames from KL-0 toward KL-4.

After this, the source image $x_S$ with the predicted noise $\hat{n}$ as the latent codes $c$ and morphing flow field $\varphi_\eta$ are fed to the Spatial Transform Network (STN) \cite{STN}. STN is a specialized type of neural network architecture that is designed to perform spatial transformations on input data. Here, we use the bilinear interpolate and sample pixel values from the latent codes $c$ on the generated grid, mapping them to the output image $x_S(\varphi_{\eta})$ based on the morphing flow field $\varphi$.

Finally, under the supervision of the pre-trained KL-2 vs. KL-3 classifier for $x_S(\varphi_{\eta=0.5})$ and $x_S(\varphi_{\eta=0.75})$, five continuous morphed images (i.e., $x_S(\varphi_{\eta=0})$, $x_S(\varphi_{\eta=0.25})$, $x_S(\varphi_{\eta=0.5})$, $x_S(\varphi_{\eta=0.75})$, and $x_S(\varphi_{\eta=1})$) are synthesized. For more intuition, Algorithm \ref{algorithm1} describes the training process of the proposed approach, with different constraint losses detailed in the following section.

\begin{algorithm}
\caption{Learning algorithm of the proposed approach}
\hspace*{\algorithmicindent} \textbf{Input:} $x_S, x_T, t$, and all initial parameters\\
\hspace*{\algorithmicindent} \textbf{Output:} $\varphi, x_S(\varphi_{\eta})$, and learned parameters\\
\hspace*{\algorithmicindent} \textbf{repeat}\\
\hspace*{\algorithmicindent} \hspace*{\algorithmicindent} \textbf{Get} $n,  x_t \leftarrow noise(x_S, t)$\\
\hspace*{\algorithmicindent} \hspace*{\algorithmicindent} \textbf{Get} $\hat{n} \leftarrow denoise(x_S, x_T, x_t, t)$\\
\hspace*{\algorithmicindent} \hspace*{\algorithmicindent} \textbf{Compute} $\mathcal{L}_{diff}(n, \hat{n})$ \Comment{Eq. \ref{diff_classical_loss}}\\
\hspace*{\algorithmicindent} \hspace*{\algorithmicindent} \textbf{Get} $\varphi_{\eta} \leftarrow morph(x_S, \hat{n})$\\
\hspace*{\algorithmicindent} \hspace*{\algorithmicindent} \textbf{Get} $x_S(\varphi_{\eta}) \leftarrow {STN}(x_S, \varphi_{\eta}, \hat{n})$\\
\hspace*{\algorithmicindent} \hspace*{\algorithmicindent} \textbf{Compute} $\mathcal{L}_{mph}(x_S(\varphi_{\eta = 1}), x_T)$ \Comment{Eq. \ref{def_loss}}\\
\hspace*{\algorithmicindent} \hspace*{\algorithmicindent} \textbf{Compute} $\mathcal{L}_{sup}(x_S(\varphi_{\eta = 0.5}), x_S(\varphi_{\eta = 0.75}))$ \Comment{Eq. \ref{sup_loss}}\\
\hspace*{\algorithmicindent} \hspace*{\algorithmicindent} \textbf{Update} Parameters of the model\\
\hspace*{\algorithmicindent} \textbf{until} \emph{convergence}\\
\label{algorithm1}
\end{algorithm}

\subsection{Hybrid loss strategy}
\label{hybrid_loss}
In the diffusion module, the diffusion loss $\mathcal{L}_{diff}$ is defined as shown in Eq. \ref{diff_classical_loss}. In the morphing module, drawing from the energy function used in traditional image registration, the morphing loss $J_{mph}$ consists of Normalized Cross-Correlation (NCC) loss and Image Gradient (IG) loss. NCC calculates the correlation between two vectors of the same dimension, with values ranging from $[-1, 1]$. A value of -1 indicates no correlation between the vectors, while a value of 1 indicates perfect correlation. The NCC loss $\mathcal{L}_{NCC}$ is defined as follows:

\begin{equation}
\mathcal{L}_{NCC} = 1 - \frac{\sum_{i,j} (A \ast B)^2}{\sqrt{\sum_{i,j} A^2 \sum_{i,j} B^2}}
\end{equation}
where $A$ and $B$ represent the input and target images, respectively. $\ast$ denotes the cross-correlation operation. $i$ and $j$ are indices of the image.

On the other hand, the IG loss calculates the mean squared Euclidean distance between the gradient channels of the reference image and the target one. It encourages the two images to have similar gradient patterns, which can help in enhancing fine details or maintaining stylistic consistency between images. The IG loss $\mathcal{L}_{IG}$ is defined as follows:

\begin{equation}
\mathcal{L}_{IG} = \sum_{i,j} \left| \frac{\partial A}{\partial x} - \frac{\partial B}{\partial x} \right| + \left| \frac{\partial A}{\partial y} - \frac{\partial B}{\partial y} \right|
\end{equation}
where $\frac{\partial A}{\partial x}$, $\frac{\partial B}{\partial x}$, and $\frac{\partial A}{\partial y}$, $\frac{\partial B}{\partial y}$ are the partial derivatives of the images $A$ and $B$ with respect to the $x$-axis (horizontal gradient) and $y$-axis (vertical gradient), respectively.

The morphing loss $J_{mph}$ is defined as:

\begin{equation}
\mathcal{L}_ {mph} = \mathcal{L}_{NCC}(x_S(  \varphi_{\eta=1}  ),x_T) + \mathcal{L}_{IG}(x_S(\varphi_{\eta=1}), x_T)
\label{def_loss}
\end{equation}

As KL grades possess a semi-quantitative nature and cannot be strictly divided into continuous intervals according to a uniform distribution, we here introduce a pre-trained classifier as a supervision module, which aims to enhance the accuracy of the KL grade assignments for intermediate frames. Since the literature often considers the label data for KL-1 patients to be questionable, these labels may involve significant uncertainties. Therefore, we focused on the supervision for the intermediate frames of KL-2 and KL-3, corresponding to the $x_S(\varphi_{\eta=0.5})$ and $x_S(\varphi_{\eta=0.75})$. To do this, the Stat-Of-The-Art (SOTA) classification model proposed in \cite{zhe_ViT} was employed. The supervision loss $\mathcal{L}_ {sup}$ is calculated as:

\begin{equation}
\mathcal{L}_ {sup} = \mathcal{L}_ {CE}(x_S(  \varphi_{\eta=0.5} ))+  \mathcal{L}_ {CE}(x_S(  \varphi_{\eta=0.75} ))
\label{sup_loss}
\end{equation}

Finally, given that the diffusion loss $\mathcal{L}_{diff}$ is the basic function of the global approach, a weight of 1 was assigned. The proposed hybrid loss function $\mathcal{L}_{hybrid}$ in this study is defined as:

\begin{equation}
\label{hybird}
\mathcal{L}_{hybrid} = \mathcal{L}_{diff} + \lambda_1 \mathcal{L}_{mph} + \lambda_2 \mathcal{L}_{sup}
\end{equation}
where the hyper-parameters $\lambda_1$ and $\lambda_2$ serve to weigh and balance the aforementioned loss functions more effectively. The influence of these parameters on performance will be examined in Section \ref{hyperparameters}.

\section{Experimental settings}
This section provides an overview of the experimental data, the preprocessing steps undertaken, and the specifics of the experimental procedures.

\subsection{Employed knee database}
The OsteoArthritis Initiative (OAI) \cite{OAI} represents a significant and rich source of data for researchers investigating KOA and related conditions. By analyzing data from 4,796 participants aged between 45 and 79 years over 96 months, the initiative provides a comprehensive longitudinal dataset. Each participant underwent nine follow-up examinations, allowing for detailed tracking of the disease's progression and the identification of potential risk factors associated with KOA development or progression.

\subsection{Data preprocessing}
\label{data_preprocessing}
\begin{figure}[htbp]
\centering
\subfigure[A standard knee plain radiograph]{
\label{plainXray}
\begin{minipage}[t]{0.27\textwidth}
\centering
\includegraphics[width=1\textwidth]{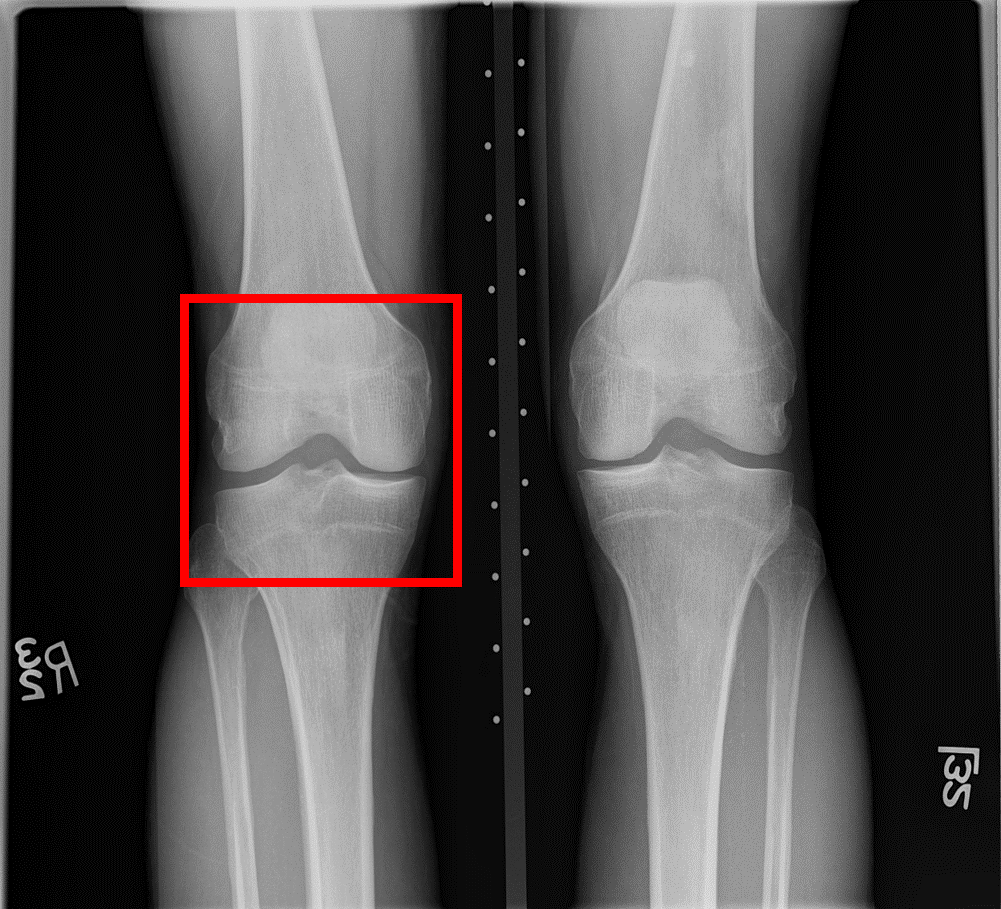}
\end{minipage}
}
\subfigure[An identified knee joint]{
\label{kneeROI}
\begin{minipage}[t]{0.256\textwidth}
\centering
\includegraphics[width=0.96\textwidth]{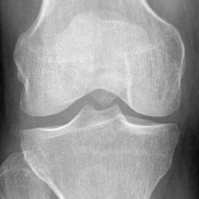}
\end{minipage}
}
\caption{A standard knee plain radiograph from the database and an identified knee joint highlighted in a red box \ref{plainXray}. An identified knee joint \ref{kneeROI}.}
\label{detectionknee}
\end{figure}

The knee joints employed in this study are derived from \cite{chen}. As shown in Fig. \ref{detectionknee}, the knee joint (Fig. \ref{kneeROI}) was identified from the plain radiograph (Fig. \ref{plainXray}) through the YOLOv2 learning model \cite{yolov2}, with images sized  299 $\times$ 299, which were resized to 224$\times$224 pixels, and image intensity was normalized to [-1, 1]. As a result of the preprocessing steps, 60 pairs of KL-0 and KL-4 images were collected from the baseline, 24-month, 36-month, 48-month, 72-month, and 96-month datasets. Each X-ray image pair was obtained from the same patient, ensuring consistency in the temporal progression of KOA within the dataset.

\subsection{Experimental details}
The weights of all modules (denoise network, registration network, and STN) were initialized using the Kaiming procedure \cite{kaiming}. To enhance the model's generalization ability, the diffusion steps $T$ are randomized in each epoch, with a maximum value set to 2,000. The Adam optimizer method \cite{adam} was utilized for training over 500 epochs. The training configuration included a mini-batch size of 16, with the denoise network using a learning rate of 2e-04, and both the registration network and STN using a learning rate of 1e-05. We implemented our approach using PyTorch v1.8.1 \cite{pytorch} on Nvidia H100 graphics cards with 80 GB of memory.

\section{Results and discussion}
This section presents and analyzes the experimental results.

\subsection{Selection of the hyper-parameters}
\label{hyperparameters}
\begin{figure}[htbp]
\centering
\includegraphics[width=0.55\textwidth]{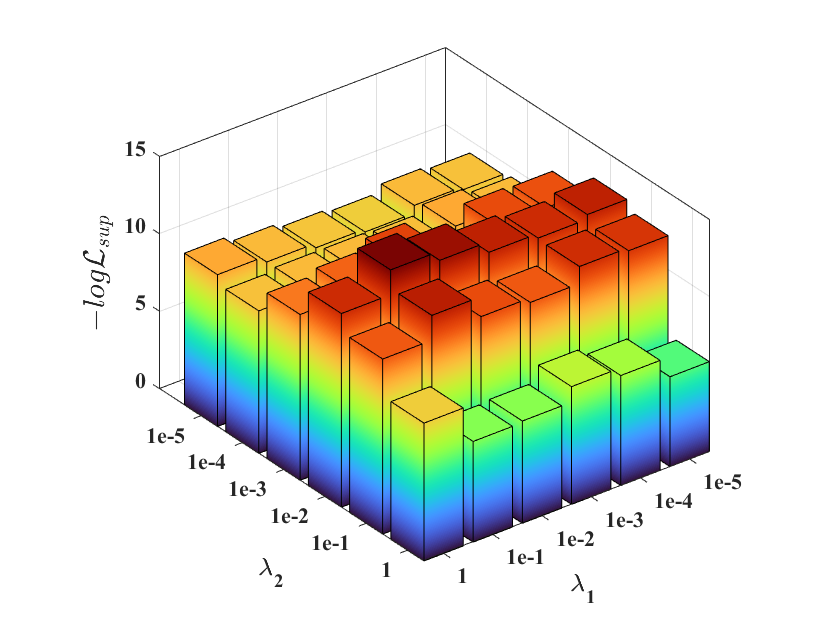}
\caption{Supervision performance metrics obtained using different values of $\lambda_1$ and $\lambda_2$ for the synthesized intermediate frames $x_S(\varphi_{\eta=0.5})$ and $x_S(\varphi_{\eta=0.75})$.}
\label{3d}
\end{figure}

To ensure the effectiveness of generating intermediate frames in advance, the weight hyper-parameters ($\lambda_1$ and $\lambda_2$) in the proposed hybrid loss function (Eq. \ref{hybird}) were tuned via a small grid search over combinations of $\lambda_1 \in [0,1]$ and $\lambda_2 \in [0,1]$. By leveraging the supervision's feedback, adjustments to the hyper-parameters $\lambda_1$ and $\lambda_2$ were made. To do this, we calculated the supervision loss $\mathcal{L}_ {sup}$ for the generated intermediate frames $x_S(\varphi_{\eta=0.5})$ and $x_S(\varphi_{\eta=0.75})$. As can be seen in Fig. \ref{3d}, the z-axis, representing the negative log of the loss function, indicates the model performance, with higher bars suggesting better performance (i.e., lower loss). After five rounds of testing, the experimental results provided insights into the trade-offs between different hyper-parameter settings. Assigning higher values to $\lambda_1$ ensures that the intermediate frames maintain a coherent motion trajectory. However, overemphasis on $\lambda_1$ may result in neglecting fine details, reducing the accuracy of KOA characteristics. On the other hand, higher values of $\lambda_2$ compromise temporal smoothness, leading to discontinuous sequences. Ultimately, the combination of $\lambda_1 = 0.1$ and $\lambda_2 = 0.01$ was selected as the optimal weights for the hybrid loss function.

\begin{table*}
\centering
\caption{Visualization of the synthesized intermediate frames$^*$}
\label{vlisualization}
\begin{threeparttable}
\setlength{\tabcolsep}{0.4mm}
\begin{tabular}{ccccc}
\toprule
Input image & \multicolumn{3}{c}{$\stackrel{\text{\footnotesize Synthesized intermediate frames}}{\makebox[10cm]{\rightarrowfill}}$} & Input image\\
$x_S$ & $x_S(\varphi_{\eta=0.25})$ & $x_S(\varphi_{\eta=0.5})$ & $x_S(\varphi_{\eta=0.75})$ & $x_T$\\
\midrule
KL-0 &KL-1$^{**}$ &KL-2$^{**}$ &KL-3$^{**}$ &KL-4\\
\midrule
\begin{minipage}[b]{0.19\columnwidth}\centering \raisebox{-.5\height}{\includegraphics[width=\linewidth]{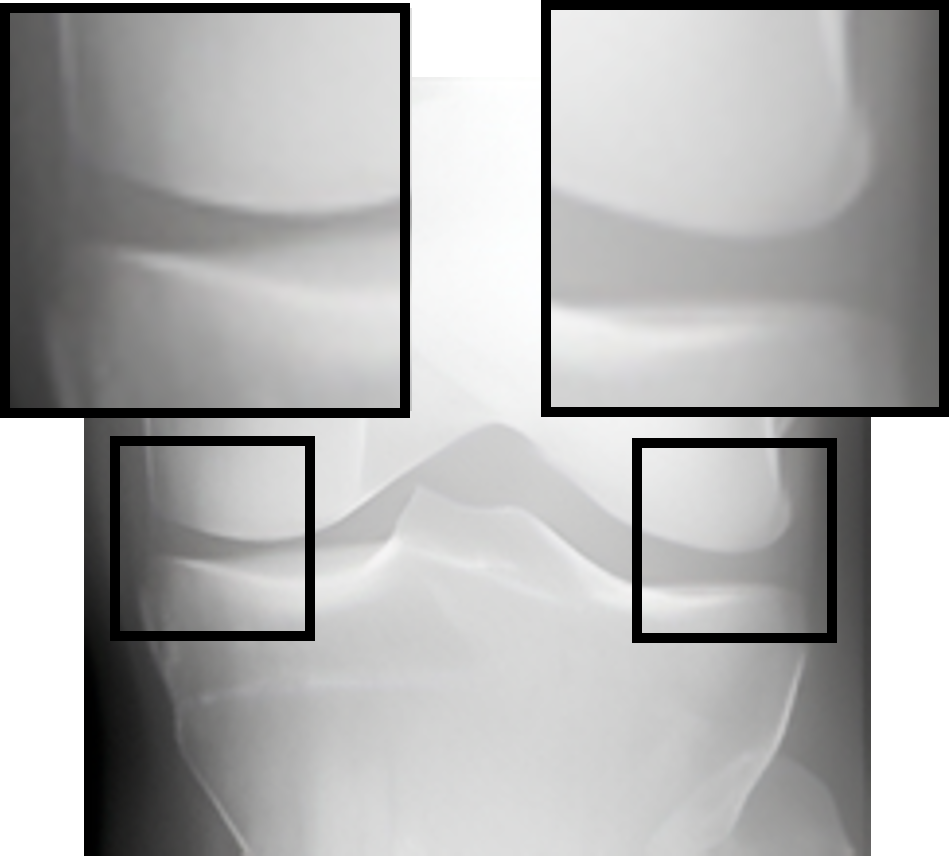}}\end{minipage}&\begin{minipage}[b]{0.19\columnwidth}\centering \raisebox{-.5\height}{\includegraphics[width=\linewidth]{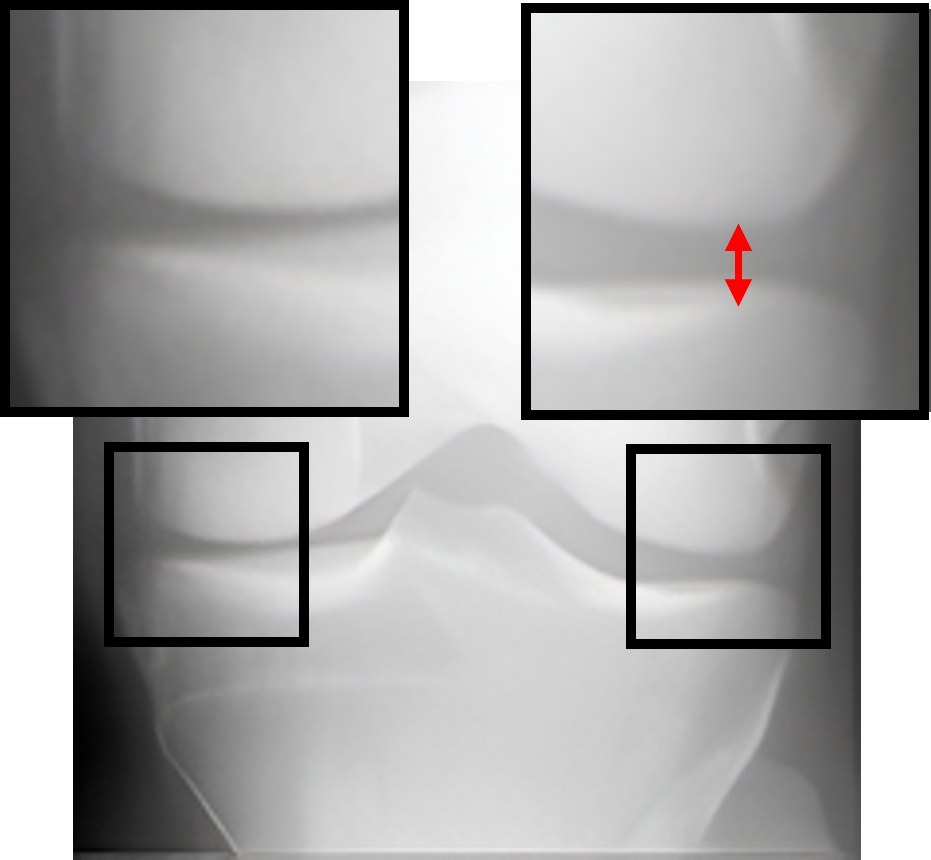}}\end{minipage}&\begin{minipage}[b]{0.19\columnwidth}\centering \raisebox{-.5\height}{\includegraphics[width=\linewidth]{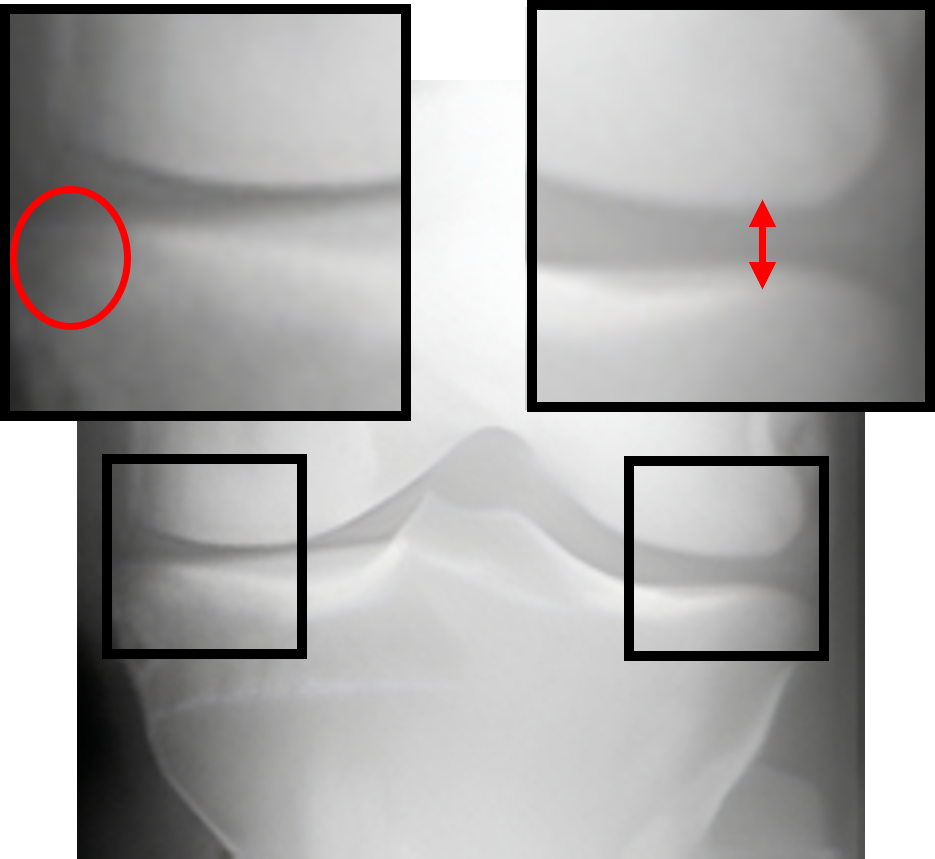}}\end{minipage}&\begin{minipage}[b]{0.19\columnwidth}\centering \raisebox{-.5\height}{\includegraphics[width=\linewidth]{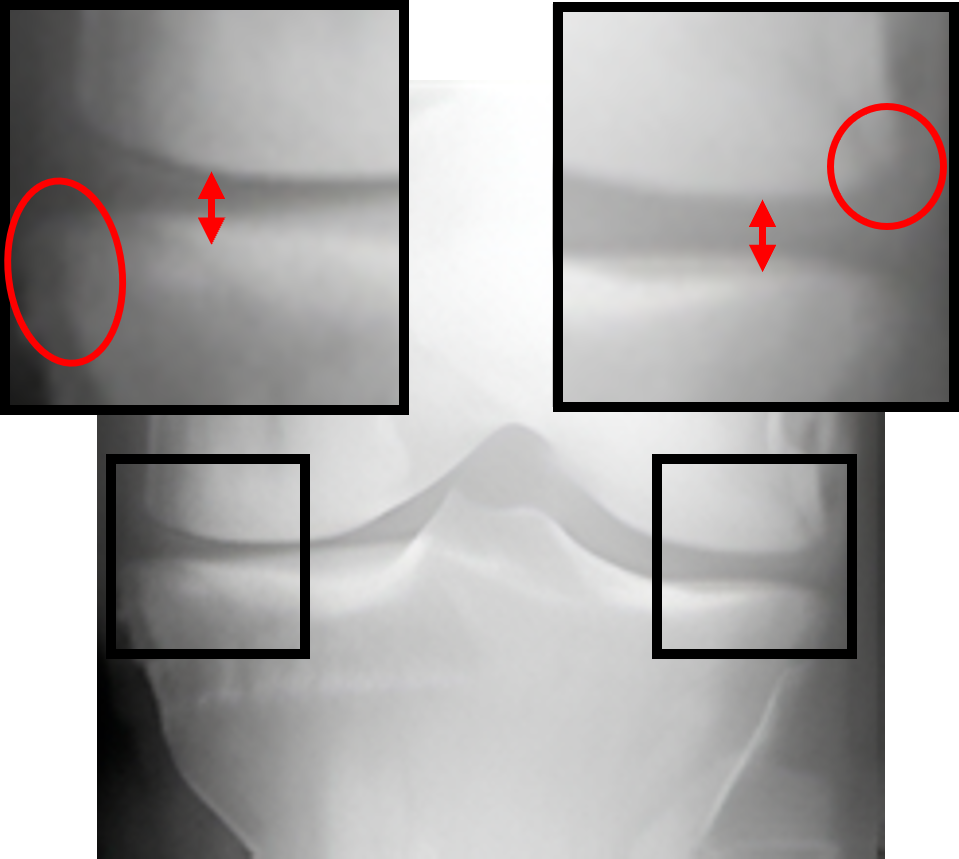}}\end{minipage}&\begin{minipage}[b]{0.19\columnwidth}\centering \raisebox{-.5\height}{\includegraphics[width=\linewidth]{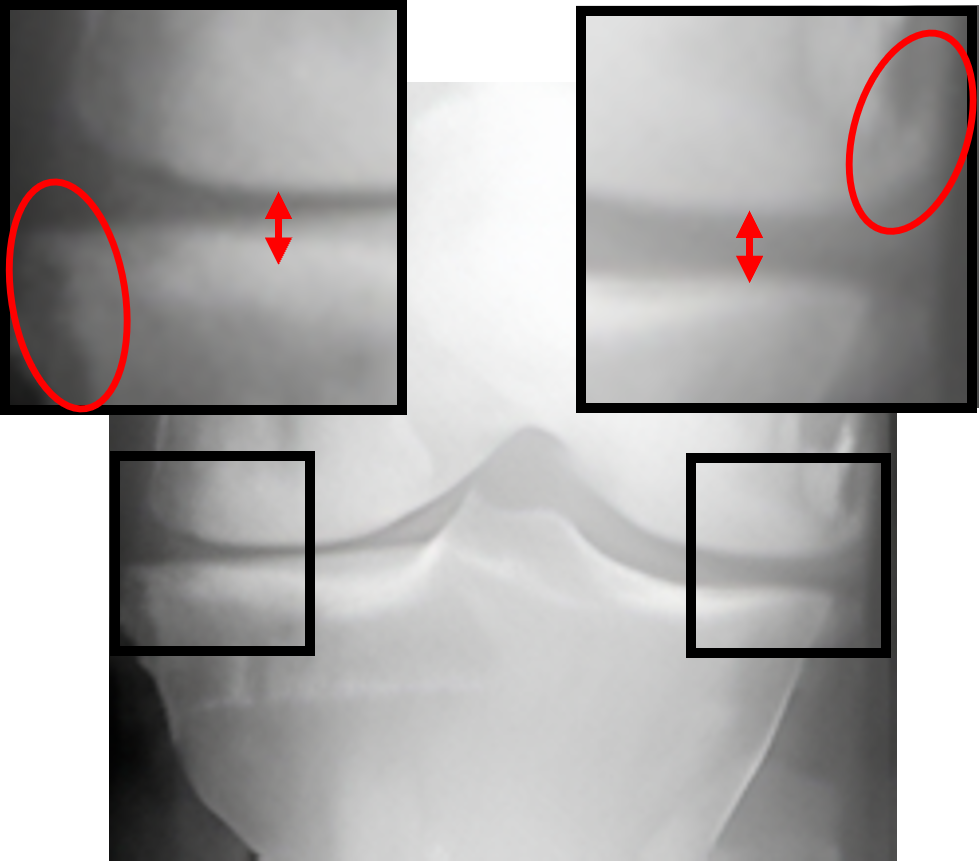}}\end{minipage}\\
\midrule
\begin{minipage}[b]{0.19\columnwidth}\centering \raisebox{-.5\height}{\includegraphics[width=\linewidth]{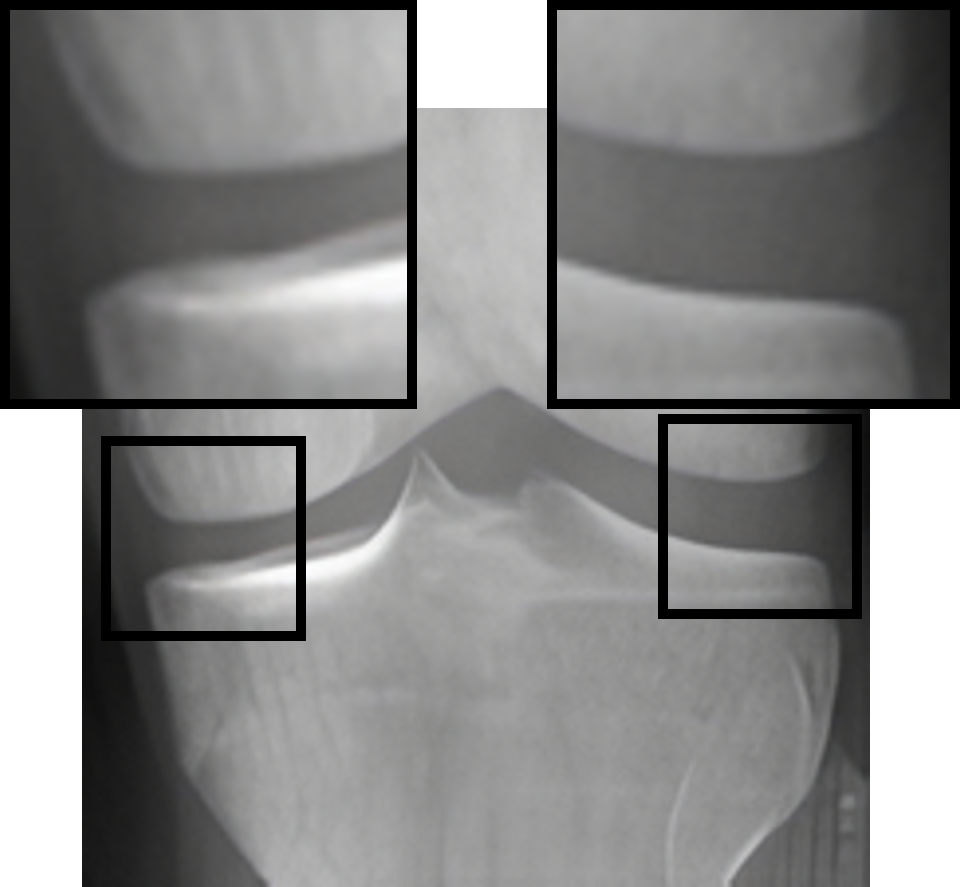}}\end{minipage}&\begin{minipage}[b]{0.19\columnwidth}\centering \raisebox{-.5\height}{\includegraphics[width=\linewidth]{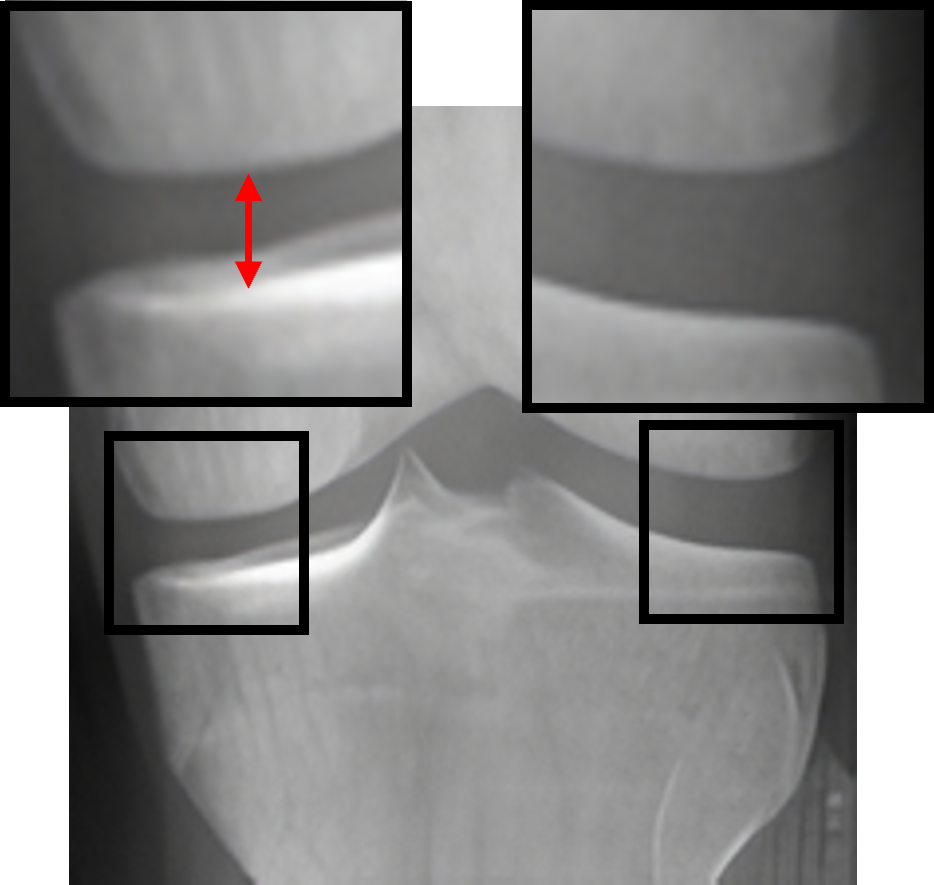}}\end{minipage}&\begin{minipage}[b]{0.19\columnwidth}\centering \raisebox{-.5\height}{\includegraphics[width=\linewidth]{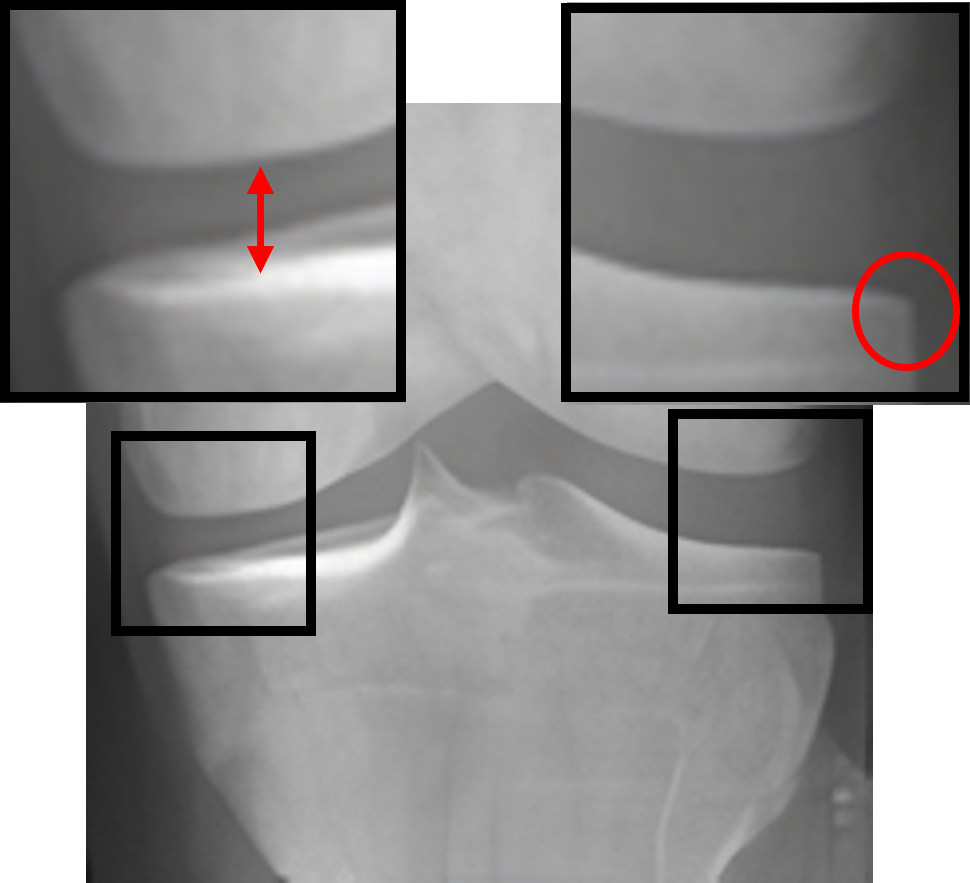}}\end{minipage}&\begin{minipage}[b]{0.19\columnwidth}\centering \raisebox{-.5\height}{\includegraphics[width=\linewidth]{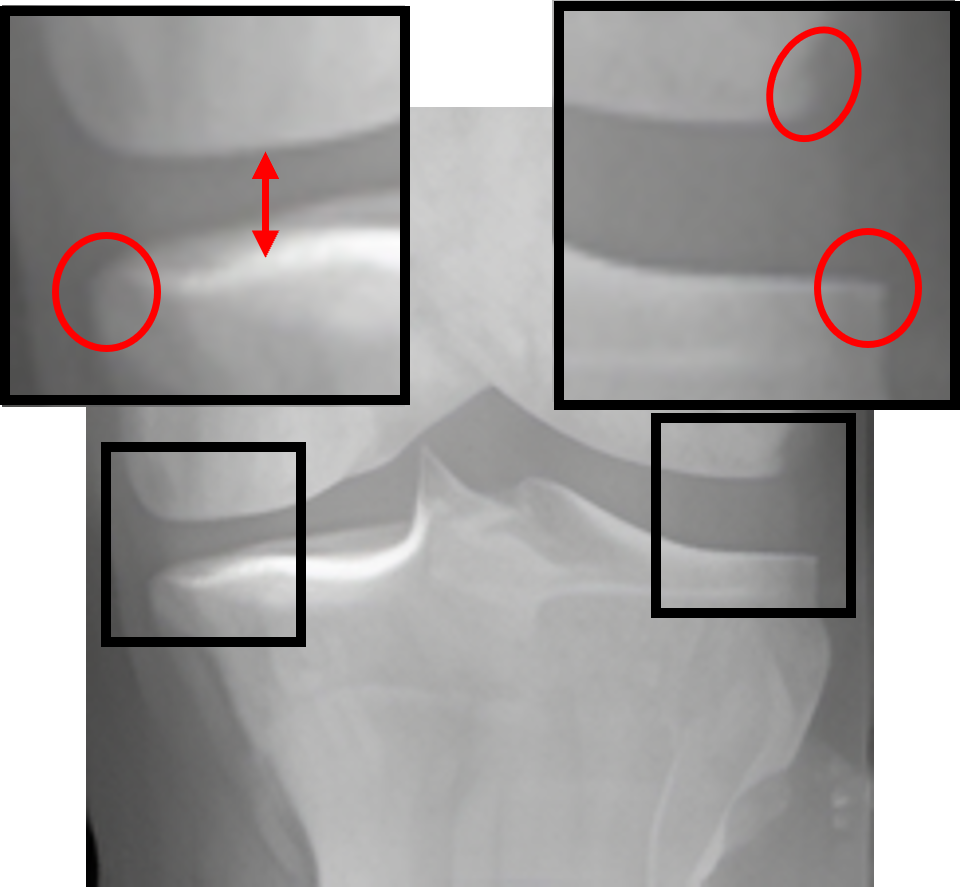}}\end{minipage}&\begin{minipage}[b]{0.19\columnwidth}\centering \raisebox{-.5\height}{\includegraphics[width=\linewidth]{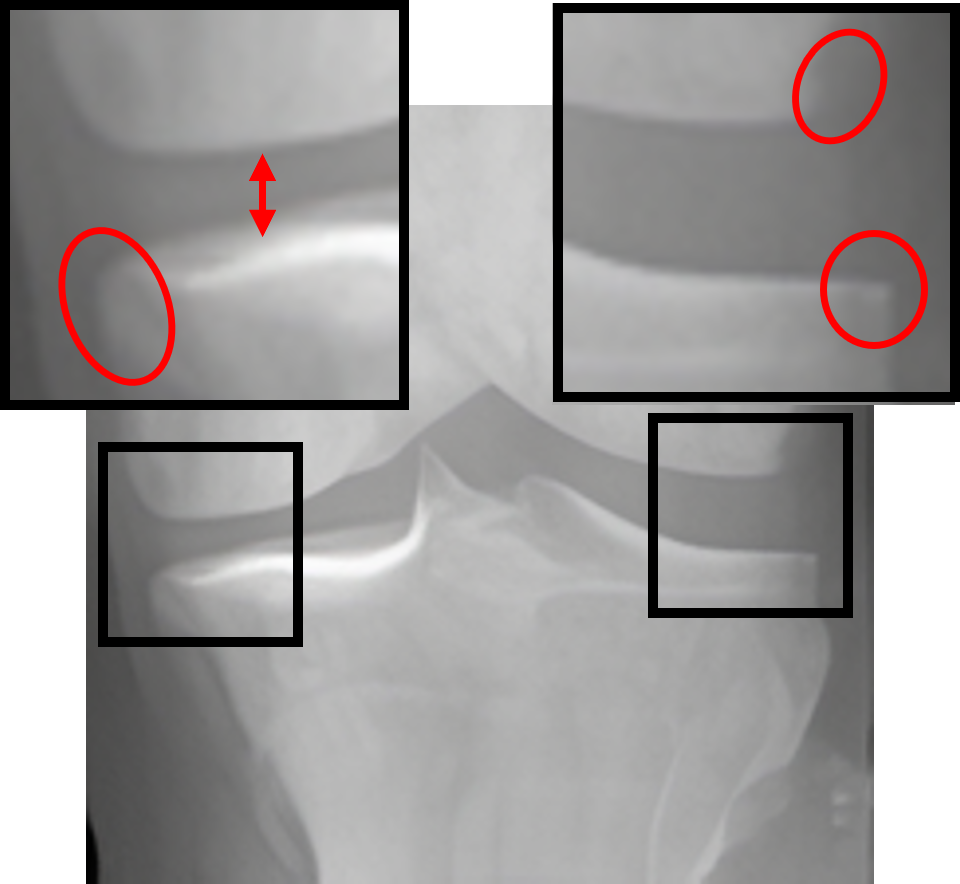}}\end{minipage}\\
\bottomrule
\end{tabular}
\begin{tablenotes}
\footnotesize
\item[$*$] The morphed areas of each frame were highlighted within black boxes, the red arrow and red circle indicate the locations of JSN and osteophyte formation, respectively. A shorter arrow indicates a smaller joint space (i.e., a narrower distance between the tibia and femur), while a larger circle indicates more osteophytes.
\item[$**$] The expected labels.
\end{tablenotes}
\end{threeparttable}
\end{table*}

\subsection{Visualization of the synthesized intermediate frames}
In Table \ref{vlisualization}, we show the source image $x_S$, the target image $x_T$, and their intermediate frames (i.e., $x_S(\varphi_{\eta=0.25})$, $x_S(\varphi_{\eta=0.5})$ and $x_S(\varphi_{\eta=0.75})$), which is crucial in understanding how the source images evolve into their registered state. The progressive alignment is quantified by the parameter $\eta$ evenly divided into 0.25, 0.5, and 0.75, which represents the extent of transformation at each stage of KOA, corresponding to KL-1, KL-2, and KL-3, respectively. As can be seen, $x_S(\varphi_{\eta=0.25})$ shows the initial stage of the morphing process where the morphs are minimal and closely resemble the source image. For $x_S(\varphi_{\eta=0.5})$, the morph is more pronounced, introducing clear signs of osteophytes and beginning to show a narrowing of the joint space. It is noteworthy that this frame is critical as it represents the halfway anchor point of the KOA temporal evolution process, showing a balanced mix of characteristics from both the source and target images. $x_S(\varphi_{\eta=0.75})$ signifies an advanced stage of morph, introducing moderate multiple osteophytes, definite narrowing of the joint space, some sclerosis, and possible deformity of bone contour.

\subsection{Comparison of different existing approaches}
\label{samples_discussion}
The evaluation of the intermediate frames was employed using the Peak Signal-to-Noise Ratio (PSNR) and the Normalized Mean Square Error (NMSE) according to the real images to quantify the quality of the generated morphed images. Specifically, PSNR is a measure used to assess the quality of reconstructed or compressed images compared to their original versions. Higher PSNR values typically signify improved reconstruction quality, as they indicate lower deviation from the original image. The formula for PSNR is:

\begin{equation}
\text{PSNR} = 10 \cdot \log_{10} \left( \frac{\text{MAX}_I^2}{\text{MSE}} \right)
\end{equation}
where $\text{MAX}_I$ is the maximum possible pixel value of the image. $\text{MSE}$ stands for the Mean Squared Error, which calculates the average of the squared differences between the real and synthesized images:

\begin{equation}
\text{MSE} = \frac{1}{mn} \sum_{i=1}^{m} \sum_{j=1}^{n} ( I(i, j) - K(i, j) )^2
\end{equation}
where $I(i,j)$ and $K(i,j)$ are the pixel values at position $(i,j)$ in the real and synthesized images, respectively. $m$ and $n$ are the width and height of the image.

On the other hand, NMSE is a measure of the quality of an estimator. It normalizes the MSE by the variance of the actual values, providing a scale-independent accuracy measure. An NMSE value close to 0 indicates a model with minimal error compared to the variance of the data. The formula for NMSE is:

\begin{equation}
\text{NMSE} = \frac{\sum_{i=1}^{n} (y_i - \hat{y}_i)^2}{\sum_{i=1}^{n} (y_i - \bar{y})^2}
\end{equation}
where $y_i$ denotes the true values, $\hat{y}_i$ signifies the predicted values, $\bar{y}$ is the average of the true values, $n$ represents the total number of observations.

\begin{figure}[htbp]
\centering
\subfigure[]{
\begin{minipage}[t]{0.32\textwidth}
\label{PSNRR}
\centering
\includegraphics[width=1\textwidth]{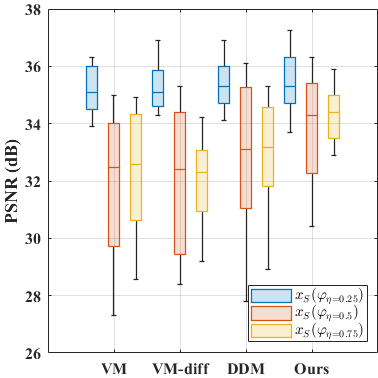}
\end{minipage}
}
\subfigure[]{
\begin{minipage}[t]{0.32\textwidth}
\label{NMSEE}
\centering
\includegraphics[width=1\textwidth]{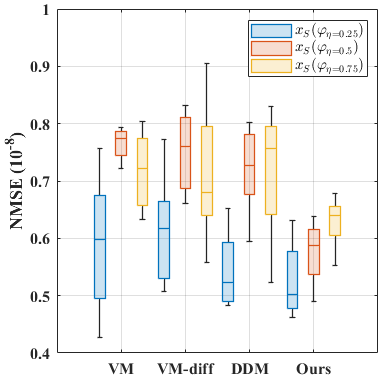}
\end{minipage}
}
\caption{The box plots visualize the different performance of the selected approaches using PSNR \ref{PSNRR} and NMSE \ref{NMSEE} for the synthesized intermediate frames $x_S(\varphi_{\eta=0.25})$, $x_S(\varphi_{\eta=0.5})$, and $x_S(\varphi_{\eta=0.75})$.}
\label{boxplots}
\end{figure}

We compared the performance of the proposed method with existing approaches, including VoxelMorph (VM) \cite{ VoxelMorph}, VoxelMorph-diff (VM-diff) \cite{VoxelMorph-diff}, and Diffusion Deformable Model (DDM) \cite{DDM}. The evaluation focused on the synthesized intermediate frames (i.e., $x_S(\varphi_{\eta=0.25})$, $x_S(\varphi_{\eta=0.5})$, and $x_S(\varphi_{\eta=0.75})$) by comparing them with the corresponding real images using two key metrics PSNR and NMSE. After five rounds of testing, the box plots illustrate the statistical distribution of these metrics. As can be seen in Fig. \ref{boxplots}, VM and VM-diff have a wider spread of PSNR values, with more variation in the results. The DDM is more consistent but generally yields lower median PSNR values than our proposed method. Especially with $x_S(\varphi_{\eta=0.5})$, our method shows a higher median PSNR and a tighter interquartile range, indicating better and more consistent image quality. On the other hand, our proposed method has a consistently lower NMSE across all three cases (i.e., $x_S(\varphi_{\eta=0.25})$, $x_S(\varphi_{\eta=0.5})$, and $x_S(\varphi_{\eta=0.75})$), which suggests a superior estimation accuracy over the compared methods. This is not only evident from the lower median values but also from the tighter interquartile ranges, implying a more reliable performance regardless of the underlying complexity characterized by $\varphi_{\eta}$. The DDM has higher variability in NMSE scores, especially for $x_S(\varphi_{\eta=0.25})$ and $x_S(\varphi_{\eta=0.5})$. The medians do not show a large difference between $x_S(\varphi_{\eta=0.25})$ and $x_S(\varphi_{\eta=0.5})$ across the groups, but there is a noticeable improvement for $x_S(\varphi_{\eta=0.75})$ using our method. Overall, the box plots indicate that our proposed method not only achieves higher PSNR values, suggesting better image quality and fidelity, but also minimizes NMSE, implying more accurate predictions of intermediate frames. These results demonstrate the efficacy of our approach in synthesizing high-quality and accurate intermediate frames for KOA.

\subsection{Generalization validation of the proposed approach}
To verify the generalization capability of the proposed approach, during the testing phase, only patient samples with KL-0 data are used as source images $x_S$ to synthesize continuous frames until the last frame $x_S(\varphi_{\eta=1})$. To achieve this, once the model converges, the average of the morphing fields $\varphi$ from the last five iterations is calculated to obtain a unified morphing field $\bar{\varphi}$. Subsequently, the model weights are frozen to maintain this state.  The synthesized frames are then evaluated using a multi-faceted approach, which aims to ensure the reliability and applicability of the synthesized frames in both clinical and research settings.

\subsubsection{Data augmentation effects}
To ensure that the experimental results more sensitively reflect the impact of the synthesized frames, the evaluation of the data augmentation effects was conducted via binary classification tasks. According to each classification task, we divided the corresponding dataset into a training set and validation set with the radio of 7:3. Then, the corresponding synthesized intermediate frames were utilized as augmented data into the training set while fixing the validation set. As presented in Table \ref{ablation}, the selected evaluation SOTA models are from \cite{tiuplin} and \cite{zhe_ViT} using the respective configurations (i.e., same number of training epochs, same learning rate, etc.). As can be seen, the incorporation of synthesized images leads to a general improvement in accuracy for all classification tasks across both evaluation models. More specifically, for Tiulpin et al. \cite{tiuplin}, the improvements range from a modest 0.27$\%$ for the KL-1 vs KL-4 task to a more substantial 3.84 $\%$ for the KL-1 vs KL-2 task. Similarly, for Wang et al. \cite{zhe_ViT}, the enhancements in accuracy are also noticeable, with the KL-0 vs KL-2 task seeing a 2.54$\%$ increase and the KL-1 vs KL-2 task experiencing a 3.29 $\%$ improvement. These figures highlight a trend where the more significant enhancements are observed in classifications involving KL-2 grade. The KL-2 represents a critical grade where KOA characteristics begin to be unequivocally observed, albeit in their nascent stages. This makes the classification between the very early stages (i.e., KL-0, KL-1, and KL-2) particularly sensitive to the augmentation of data volume, as the additional nuances captured through data augmentation can significantly aid in the identification of early degenerative changes. On the other hand, for the higher grades (i.e., KL-3 and KL-4), where KOA characteristics are more pronounced and easier to be identified even with smaller datasets, the incremental improvements achieved through data augmentation are less dramatic.

\begin{table}[htbp]
\centering
\caption{Analysis of data augmentation effects}
\begin{threeparttable}
\setlength{\tabcolsep}{2.5mm}
\begin{tabular}{lcccc}
\toprule
\multirow{3.5}{*}{Model} & \multirow{3.5}{*}{Task} & \multicolumn{2}{c}{Acc ($\%$)} & \multirow{3.5}{*}{Diff ($\%$)}\\
\cmidrule(lr){3-4}
 &  & \multirow{2}{*}{\makecell[c]{Original \\ training set}} & \multirow{2}{*}{\makecell[c]{Augmented \\ training set$^*$}} & \\
 &&&&\\
\midrule
\multirow{9}{*}{Tiulpin et al. \cite{tiuplin}} & KL-0 vs KL-1 & 65.88 & 68.11 & 2.23 $\uparrow$\\ 
&KL-0 vs KL-2 & 87.33 & 89.13 & 1.80 $\uparrow$\\ 
&KL-0 vs KL-3 & 96.29  & 97.01 & 0.72 $\uparrow$\\
&KL-1 vs KL-2 & 70.33 & 74.17 & 3.84 $\uparrow$\\
&KL-1 vs KL-3 & 93.17 & 95.87 & 2.70 $\uparrow$\\
&KL-1 vs KL-4 & 97.22 & 97.49 & 0.27 $\uparrow$\\
&KL-2 vs KL-3 & 94.87 & 95.91 & 1.04 $\uparrow$\\
&KL-2 vs KL-4 & 96.66 & 97.23 & 0.57 $\uparrow$\\
&KL-3 vs KL-4 & 95.11 & 95.99 & 0.88 $\uparrow$\\
\midrule
\multirow{9}{*}{Wang et al. \cite{zhe_ViT}} & KL-0 vs KL-1 & 69.08 & 70.21 & 1.13 $\uparrow$\\ 
&KL-0 vs KL-2 & 89.80 & 92.34 & 2.54 $\uparrow$\\ 
&KL-0 vs KL-3 &  98.75 & 99.11 & 0.36 $\uparrow$\\
&KL-1 vs KL-2 & 77.02 & 80.31 & 3.29 $\uparrow$\\
&KL-1 vs KL-3 & 96.84 & 97.93 & 1.09 $\uparrow$\\
&KL-1 vs KL-4 & 97.47 & 97.71 & 0.24 $\uparrow$\\
&KL-2 vs KL-3 & 95.53 & 96.59 & 1.06 $\uparrow$\\
&KL-2 vs KL-4 & 97.10 & 97.99 & 0.89 $\uparrow$\\
&KL-3 vs KL-4 & 95.88 & 96.25 & 0.37 $\uparrow$\\
\bottomrule
\end{tabular}
\begin{tablenotes}
\footnotesize
\item[$*$] The augmented training set consists of the original training data and the synthesized intermediate frames.
\end{tablenotes}
\end{threeparttable}
\label{ablation}
\end{table} 

\subsubsection{Visualization analysis}
To visualize the differences before and after incorporating synthesized frames during training, t-distributed Stochastic Neighbor Embedding (t-SNE) \cite{tsne} was applied to reveal the high-dimensional feature space on the validation set and to highlight the generalization differences of the model. For better clarity, we focus on the impact of synthesized frames specifically on the KL-0 vs. KL-1 and KL-1 vs. KL-2 classification tasks, as they demonstrate the most significant performance improvements. As shown in Fig. \ref{tnse}, training on a dataset augmented with synthesized data has significantly improved the model's ability to discern and differentiate between data points within the same class. Specifically, the reduction in intra-class distance (i.e., distance between data points within the same class) indicates that data points within each class are closer together in the feature space, resulting in denser and more defined clusters. Simultaneously, the preservation of inter-class distance (i.e., centre points of two classes) ensures that different classes remain distinct, preventing overlap and confusion. By achieving this balance, the augmented training set greatly improves the model's generalization capabilities.

\begin{table}[htbp]
\centering
\caption{t-SNE scatter plots on the fixed validation set}
\label{tnse}
\setlength{\tabcolsep}{1.3mm}
\begin{tabular}{ccc}
\toprule
& Using original training set & Using augmented training set\\
\midrule
\makecell[c]{KL-0 \\vs \\KL-1}& \begin{minipage}[b]{0.25\columnwidth}\centering \raisebox{-.5\height}{\includegraphics[width=\linewidth]{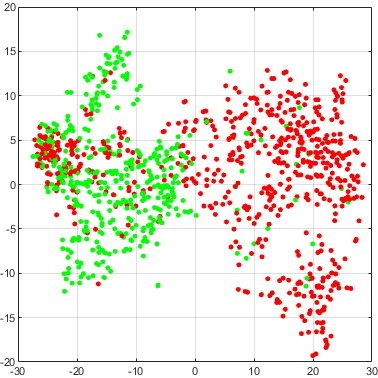}}\end{minipage}&\begin{minipage}[b]{0.25\columnwidth}\centering \raisebox{-.5\height}{\includegraphics[width=\linewidth]{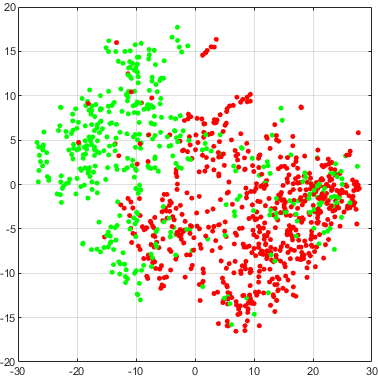}}\end{minipage}\\
&(a) & (b)\\
\midrule
\makecell[c]{KL-1 \\vs \\KL-2}& \begin{minipage}[b]{0.25\columnwidth}\centering \raisebox{-.5\height}{\includegraphics[width=\linewidth]{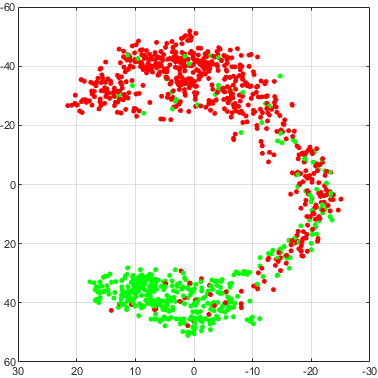}}\end{minipage}&\begin{minipage}[b]{0.25\columnwidth}\centering \raisebox{-.5\height}{\includegraphics[width=\linewidth]{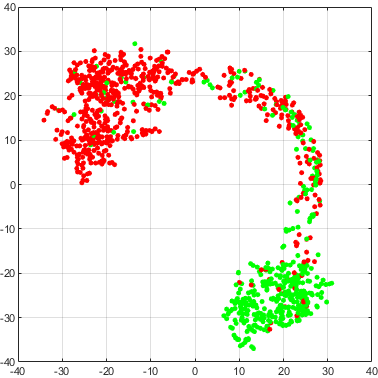}}\end{minipage}\\
&(c) & (d)\\
\bottomrule
\end{tabular}
\end{table}

\subsubsection{Expert assessment}
\label{expert_ass}
Interventional Radiologists (IR) played a critical role in the evaluation process, particularly for the ambiguous KL-1 grade. To evaluate the accuracy and quality of the synthesized frames, three radiologists were involved. They provided expert assessments on KOA diagnosis based on the diagnostic criteria and their rich experience. As shown in Table \ref{e_a}, the quality (QTY) of the synthesized intermediate frames received an average score of 3.9 out of 5.0, indicating that the quality aligns with expert expectations. Regarding accuracy (Acc), which measures how well the synthesized frames match the expected labels, the overall average accuracy was a promising 89$\%$, except for the KL-1 frames. On the other hand, an assessment of $x_S(\varphi_{\eta=0.25})$ frames reveals greater variability. The average accuracy drops to 62$\%$, indicating that a substantial proportion of these frames did not correctly match the KL-1 label, as evidenced by the lowest accuracy score of only 50$\%$, which could be attributed to the unclear definition of KL-1, leading to differing interpretations among the experts. Furthermore, the discrepancy in accuracy scores among the experts, ranging from as high as 74$\%$ to as low as 50$\%$, may depend on their experience with extremely early-stage KOA, underscoring the subjective element inherent in medical image assessment. In contrast, the accuracy rates for $x_S(\varphi_{\eta=0.5})$, $x_S(\varphi_{\eta=0.75})$, and $x_S(\varphi_{\eta=1})$ frames are consistently high, often exceeding 85$\%$, and remain relatively stable. 

\begin{table}[htbp]
\centering
\caption{Analysis of the experts' assessment}
\begin{threeparttable}
\setlength{\tabcolsep}{2.5mm}
\begin{tabular}{ccccccccc} 
\toprule
\multirow{6}{*}{IR} & \multicolumn{8}{c}{Generated frames from KL-0}\\
&\multicolumn{8}{c}{$\stackrel{}{\makebox[8.5cm]{\rightarrowfill}}$}\\
& \multicolumn{2}{c}{KL-1} & \multicolumn{2}{c}{KL-2} & \multicolumn{2}{c}{KL-3} & \multicolumn{2}{c}{KL-4}\\
&  \multicolumn{2}{c}{$x_S(\varphi_{\eta=0.25})$} & \multicolumn{2}{c}{$x_S(\varphi_{\eta=0.5})$}  &  \multicolumn{2}{c}{$x_S(\varphi_{\eta=0.75})$} &  \multicolumn{2}{c}{$x_S(\varphi_{\eta=1})$}\\
\cmidrule(lr){2-3}
\cmidrule(lr){4-5}
\cmidrule(lr){6-7}
\cmidrule(lr){8-9}
 &QTY& Acc&QTY &Acc&QTY& Acc&QTY & Acc\\
\midrule
$\#$1 & 3.8 & 62  & 4.2 & 89 & 4.1 & 88 & 4.3 & 91\\
$\#$2 & 3.5 & 74  & 3.9 & 85 & 3.9 & 89 & 4.0 & 87\\
$\#$3 & 3.4 & 50 & 4.0 & 83 & 3.7 & 85 & 4.2 & 93\\
\midrule
Average & 3.6 & 62 & 4.0 & 86 & 3.9 &87 & 4.2 & 90 \\
\bottomrule
\end{tabular}
\begin{tablenotes}
\footnotesize
\item[] QTY: Subjective image quality rating out of 5.0.
\item[] Acc: The rate at which the labels of generated intermediate frames match the labels after expert assessment.
\end{tablenotes}
\end{threeparttable}
\label{e_a}
\end{table}

Overall, the proposed approach demonstrates strong generalization capability. The synthesized frames are not only technically sound but also clinically relevant, underscoring the potential of the approach to improve diagnostic accuracy and support clinical decision-making.

\subsection{Discussion}
In this study, we introduced a novel Diffusion-based Morphing Model (DMM) to synthesize intermediate frames between KL-0 and KL-4 knee radiographs. Leveraging the DDPM, our approach uniquely integrates diffusion and morphing processes to simulate the temporal evolution of KOA. The proposed hybrid loss strategy, which incorporates diffusion loss, morphing loss, and supervision loss, effectively facilitated the learning of spatial morphing information between the source and target images. Experimental evaluations have demonstrated the superiority of our approach in terms of PSNR and NMSE, highlighting its efficacy in synthesizing high-quality and accurate intermediate frames. Additionally, using these synthesized frames for data augmentation has significantly enhanced the performance of KOA classification tasks. During the test phase, positive assessments from radiology experts validated the clinical relevance of our approach, further confirming the high quality and precise grading of the synthesized images. There are several points to discuss.

\subsubsection{Details of the hybrid loss}
As presented in Section \ref{hybrid_loss}, the proposed hybrid loss consists of several components, each targeting specific aspects of the synthesized images. Specifically, the morphing loss $J_{mph}$ is responsible for the morphing related to JSN and osteophytes, which is critical as their presence and severity can significantly influence the KL grade of the synthesized knee images. On the other hand, the supervision loss $J_{sup}$ involves a supervision module that supervises the degree of the morphed JSN and osteophytes to ensure they accurately correspond to the respective KL grades at the anchor points images (i.e., $x_S(\varphi_{\eta=0.5})$ and $x_S(\varphi_{\eta=0.75})$).

\subsubsection{Strengths and limitations}
This study has several notable strengths. Clinically, enhancing the classification performance for early KOA diagnosis (i.e., KL-0 vs KL-2 and KL-1 vs KL-2) is highly significant, as it can guide patients to undertake timely physical interventions and potentially delay the onset and progression of KOA symptoms \cite{ledingham2017diagnosis}. The synthesized intermediate frames are key in evaluating the effectiveness and accuracy of the morphing image registration algorithm as they provide insights into the algorithm's behaviour at different stages. By analyzing these frames, researchers and practitioners can gain a deeper understanding of the morphing process. Moreover, to enhance the clinical applicability of our approach, a multi-faceted approach was conducted to evaluate the synthesized frames. We believe such a stable and interpretable model within a Computer-Aided Diagnosis (CAD) system can significantly enhance the trust and acceptance of deep learning methods among medical practitioners for clinical use. There are also several limitations in our study. Variations in imaging systems and software across different studies result in differences in detailed imaging data, such as Dots Per Inch (DPI), pixel size, and other specifics. Consequently, our proposed approach has not been validated across multiple datasets. These discrepancies could introduce biases in the synthesized intermediate frames, potentially affecting the model's performance. Future work should focus on developing normalization or adaptation techniques to accommodate the variability in imaging data and automating hyper-parameter adjustments in the hybrid loss to broaden the applicability and enhance the efficiency of our approach.


\section{Statements}
This manuscript was prepared using data from the OAI. The views expressed in it are those of the authors and do not necessarily reflect the opinions of the OAI investigators, the National Institutes of Health (NIH), or the private funding partners.

\section{Acknowledgements}
The authors gratefully acknowledge the support of the National Institute of Arthritis and Musculoskeletal and Skin Diseases (NIAMS) of the NIH under award number K23AR084603, the Osteoarthritis Research Fund from Massachusetts General Hospital (MGH), Harvard Medical School (HMS).

\bibliographystyle{unsrt}  
\bibliography{references}

\end{document}